\documentclass[reprint, secnumarabic, amssymb, amsmath, superscriptaddress, longbibliography, aps, prl]{revtex4-1}
\usepackage{amsmath}
\usepackage{graphicx}

\usepackage{xcolor}

\begin{document}

\title{Resonant soft x-ray scattering from stripe-ordered La$_{2-x}$Ba$_x$CuO$_4$ detected by a transition edge sensor array detector}


\author{Young Il Joe}
%\thanks{}
\affiliation{National Institute of Standards and Technology, Boulder, CO, 80305, USA}

\author{Yizhi Fang}
\affiliation{Department of Physics and Materials Research Laboratory, University of Illinois, Urbana, IL, 61801, USA}

\author{Sangjun Lee}
\affiliation{Department of Physics and Materials Research Laboratory, University of Illinois, Urbana, IL, 61801, USA}

\author{Stella X. L. Sun}
\affiliation{Department of Physics and Materials Research Laboratory, University of Illinois, Urbana, IL, 61801, USA}

\author{Gilberto A. de la Pe{\~n}a}
\affiliation{Department of Physics and Materials Research Laboratory, University of Illinois, Urbana, IL, 61801, USA}

\author{William B. Doriese}
\affiliation{National Institute of Standards and Technology, Boulder, CO, 80305, USA}

\author{Kelsey M. Morgan}
\affiliation{National Institute of Standards and Technology, Boulder, CO, 80305, USA}

\author{Joseph W. Fowler}
\affiliation{National Institute of Standards and Technology, Boulder, CO, 80305, USA}

\author{Leila R. Vale}
\affiliation{National Institute of Standards and Technology, Boulder, CO, 80305, USA}

\author{Fanny Rodolakis}
\affiliation{Advanced Photon Source, Argonne National Laboratory, Argonne, IL, 60439, USA}

\author{Jessica L. McChesney}
\affiliation{Advanced Photon Source, Argonne National Laboratory, Argonne, IL, 60439, USA}

\author{Joel N. Ullom}
\affiliation{National Institute of Standards and Technology, Boulder, CO, 80305, USA}

\author{Daniel S. Swetz}
\affiliation{National Institute of Standards and Technology, Boulder, CO, 80305, USA}

\author{Peter Abbamonte}
\affiliation{Department of Physics and Materials Research Laboratory, University of Illinois, Urbana, IL, 61801, USA}


\date{\today}

\begin{abstract}

Resonant soft x-ray scattering (RSXS) is a leading probe of valence band order in materials best known for detecting charge density wave order in the copper-oxide superconductors. One of the biggest limitations on the RSXS technique is the presence of a severe fluorescence background which, like the RSXS cross section itself, is enhanced under resonant conditions. This background prevents the study of weak signals such as diffuse scattering from glassy or fluctuating order that is spread widely over momentum space. Recent advances in superconducting transition edge sensor (TES) detectors have led to major improvements in energy resolution and detection efficiency in the soft x-ray range. Here, we perform a RSXS study of stripe-ordered La$_{2-x}$Ba$_x$CuO$_4$ at the Cu $L_{3/2}$ edge (932.2 eV) using a TES detector with 1.5 eV resolution, to evaluate its utility for mitigating the fluorescence background problem. We find that, for suitable degree of detuning from the resonance, the TES rejects the fluorescence background, leading to a 5 to 10 times improvement in the statistical quality of the data compared to an equivalent, energy-integrated measurement. We conclude that a TES presents a promising approach to reducing background in RSXS studies and may lead to new discoveries in materials exhibiting valence band order that is fluctuating or glassy. 
\end{abstract}

%\keywords{}

\maketitle

\section{INTRODUCTION}
The strong interplay between charge, spin, orbital, and lattice degrees of freedom in strongly interacting quantum materials leads to a myriad of emergent phenomena and phases. In 3$d$ transition metal oxides, for example, the comparable size of oxygen covalency and on-site Coulomb repulsion, as well as the magnetic nature of partially filled 3$d$ orbitals, render all of the charge, spin, orbital, and lattice degrees of freedom equal in contriving novel physics such as prominent valence band instabilities at comparatively high temperature and colossal response to varying types of external fields. 

Resonant soft x-ray scattering (RSXS) is now widely recognized as one of the leading techniques for probing valence band order in quantum materials \cite{abbamonte2002,abbamonte2012,fink2013,comin2016}, one of its most prominent achievements being the discovery of charge density wave (CDW) order in nearly all families of copper-oxide superconductors \cite{abbamonte2005,ghiringhelli2012,comin2014,dasilva2014,dasilva2015,tabis2014,smadici2009,comin2016}. RSXS is a quasielastic scattering technique that exploits dispersion corrections to the atomic scattering factors due to the resonance between core and valence levels to achieve valence band sensitivity. The RSXS process involves all the intermediate states with a core hole and an additional electron in the valence shell, in principle measuring an electron spectral function similar to that probed in scanning tunneling microscopy (STM) experiments \cite{abbamonte2012}. 

When an x-ray beam is tuned to resonance, however, it is not only the resonant scattering that is amplified but also the photo-absorption, which is the first step in resonant scattering. The vast majority of photo-absorbed excited states decay through radiative channels and don't contribute to coherent scattering. Although radiative processes are almost isotropic, the sheer size of the photo-absorption cross section in the soft x-ray regime makes core level fluorescence the dominant signal in RSXS. This inelastic background is a serious limitation on the RSXS technique, particularly when trying to detect faint signals from glassy or fluctuating order in which the scattering structure factor is spread over momentum space in a diffuse manner. 

A potential solution to this problem is to use an energy-resolving detector. In hard x-ray experiments, fluorescence background is often eliminated by using proportional, solid state detectors that discriminate between photons with different energy. Unfortunately, this approach has not been possible in RSXS because highly efficient semiconductor detectors have not been widely available for soft x-rays. Furthermore, the fluorescence emission in soft x-ray experiments is near-threshold---typically within a few eV of the coherently scattered light---and semiconductor detectors lack the energy resolution needed to discriminate in this way. 

Transition edge sensors (TES) are highly sensitive photon detectors that can be configured to operate in energy-resolving mode \cite{Irwin2005}. TES detectors have been used successfully for cosmic microwave background surveys \cite{Irwin2005}, x-ray emission spectroscopy \cite{cheese2015,Palosaari2016,miajaavila2016}, time resolved x-ray absoprtion spectroscopy \cite{oneil2017}, hadronic atom spectroscopy \cite{HEATES2016}, gamma ray spectroscopy \cite{Bennett2012,Hatakeyama2014}, alpha particle spectrometry \cite{Croce2012}, among many other applications \cite{Fowler2017,Faverzani2016,Ullom2015,Morgan2018}. 
Thanks to recent advancements in their performance and usability, such detectors have begun to be adopted at large-scale x-ray facilities \cite{Ullom2014,Doriese2017,li2019,sainio2019,titus2017,Li2018}. 
State of the art TES array detectors offer superior energy resolution to solid state detectors (about 1.5 eV FWHM at 900 eV), excellent quantum efficiency, and solid angle coverage similar to spectrometers based on diffraction gratings \cite{Doriese2017, dean2013, dean2017, cheese2015}. Moreover, modern TES instruments comprise hundreds of such sensors, each of which acts as an independent, energy-resolving spectrometer. This significantly boosts the data collection rate by allowing parallel studies of multiple reciprocal space points.

Here, we present a study of the charge stripe order in La$_{2-x}$Ba$_x$CuO$_4$ ($x=0.125$) (LBCO) using a TES array detector. LBCO is a prototypical high-temperature superconductor whose charge order has been studied extensively with RSXS techniques using energy-integrating detectors \cite{abbamonte2005,comin2016}. Our goal is to use this material as a test case to quantify the advantages of introducing energy discrimination into such measurements, and hopefully identify the central issues that arise when using such techniques to study valence band phenomena more generally. 

%\cite{Arrigoni2004}.

\subsection{EXPERIMENTAL SETUP}
%\subsubsection{TES detector}

TESs are cryogenic energy sensors exploiting the sharp superconducting-to-normal phase transition to achieve energy resolution. TESs operate in an electrothermal steady state in contact with a cryogenic heat bath and in constant voltage bias, which puts the TES at its
superconducting transition. When an x-ray impinges on a TES, its temperature and, in turn, resistance increase through a continuous superconducting-to-normal transition. Consequently, the current through the TES temporarily drops. Shortly after, it recovers to its quiescent level aided by negative electrothermal feedback. This transient current drop is measured by a SQUID readout system through mutual inductance. The magnitude of the current pulse is roughly proportional to the energy of the x-ray.

Johnson noise and thermal fluctuation between a TES and the heat bath are the dominant limiting factors determining the energy resolution \cite{Irwin2005}. The resolution of a TES depends on its heat capacity, $C$, and is approximately given by 

\begin{equation}
\Delta E \sim \sqrt{k_\text{B} T_\text{op}^2 C / \alpha_I}
\end{equation}

\noindent
where $\alpha_I = \left ( T_\text{op} / R_\text{op} \right ) \partial R / \partial T |_{I_\text{op}}$, $T_\text{op}$ = 107 mK is the operating temperature of the TES, $R_\text{op}$ is its resistance in its quiescent state, and the derivative of the resistance curve, $R(T)$, is evaluated at $T_\text{op}$ and at the bias current, $I_\text{op}$. 

\begin{figure}
    \centering
    \includegraphics[width=0.495\textwidth]{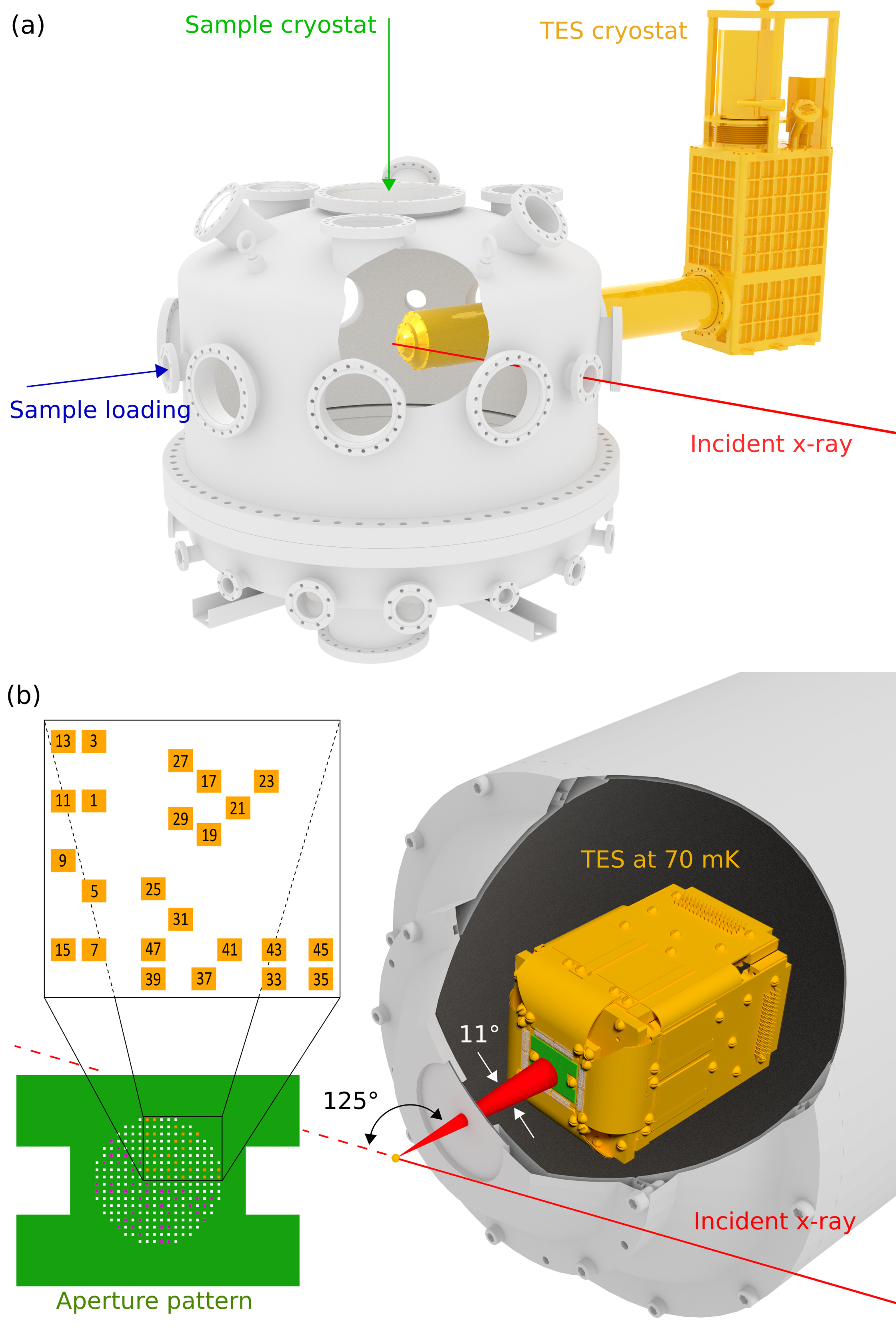}
    \caption{Schematic of the TES / RSXS scattering setup at Sector 29 at the Advanced Photon Source. (a) 
Layout of the scattering chamber (gray) showing the location of the TES detector snout and cryostat (yellow). 
The TES is connected to the chamber via a vacuum bellows and can be slid in and out of its port to change the sample distance.  In these experiments, the full detector array covered $\sim$11 degrees of scattering angle.
(b) Expanded view of the sensor head (yellow) and detector array (green). The TES consists of 240 individually addressable sensors. The data presented here were taken with a 24-element subsector of the array (orange expanded view). }
    \label{fig:rsxs_scattering_geometry}
\end{figure}

The TES detector used in these experiments was built by the Quantum Sensors Group at NIST (Boulder, CO, USA) \cite{Doriese2017,QSGweb}. The detector consists of an array of 240 sensors laid out as shown in Fig.~1(b). The exact locations of the sensors on the detector chip were chosen to allow space for wiring buses to run between them. The sensor array is maintained at a bath temperature of 70 mK via an adiabatic demagnetization refrigerator (ADR) and is mounted on the end of a long snout so it can be positioned near the sample in the sample chamber used for RSXS experiments (Fig.~1(a)).
The quantum efficiency of the TES detector is limited by transmission through a series of vacuum and radiation windows, and is about 24\% at the O $K$ edge and 57\% at the Cu L$_{3/2}$ edge \cite{Doriese2017}.

The cryogenic system requires strict management of the thermal load from every part of the device including the readout system. Thus, the number of room temperature wires into the readout chip needs to be minimized. This is achieved by implementing a time division multiplexing (TDM) scheme in which the readout system addresses multiple sensors by switching from one to another \cite{Doriese2016}. TDM results in a tradeoff between the number of sensors used and the energy resolution, since addressing more sensors results in fewer samples per channel and hence increased readout noise and arrival time errors \cite{Doriese2016}. The current experiments were done with a 24-sensor subsector of the TES, highlighted in Fig.~1(b), to illustrate the advantages of multiplexing while operating close to the optimal  resolution. All TES spectra in this study were created by binning individual photon events into histograms comprising 0.1 eV intervals.

The energy resolution was determined by measuring diffuse elastic scattering from a thin layer of gold on polished silicon, as illustrated in Fig.~2(a). The energy bandwidth of the elastic scattering is the same as that of the incident beam, which was about 0.1 eV. Since the resolution of TES is significantly larger than the elastic energy bandwidth, the observed spectrum is an excellent measure of the intrinsic resolution of the TES. The FWHM resolution varied slightly among different sensors but was approximately $\Delta E = 1.5$ eV (Fig.~2(a)).

%\subsubsection{Scattering setup}

RSXS experiments were carried out at the IEX facility in Sector 29 at the Advanced Photon Source using a two-axis reflectometer with a base sample temperature of 18 K. The LBCO crystal used had a doping level $x \approx 1/8$ with a superconducting transition temperature of  $T_c = 4$ K. The crystal was cleaved in air to obtain a fresh surface along the (0, 0, 1) crystallographic plane. The crystal was pre-oriented by x-ray diffraction (XRD) and mounted so that the reciprocal lattice vector of the charge order, which is around (0.25, 0, 1.5), lay in the scattering plane.
This allowed the CDW reflection to be optimized despite the restricted number of sample motions of the instrument. The TES detector was mounted on the RSXS chamber on a fixed port at a scattering angle of approximately $2\theta = 125^\circ$ (Fig.~1(a)). The energy response curve of the TES was measured between every pair of scan sets and used to correct for small drifts in detector calibration and gain, as described previously \cite{Doriese2017,fowler2016}. 

\subsection{RESULTS}

\begin{figure}
    \centering
    \includegraphics[width=0.495\textwidth]{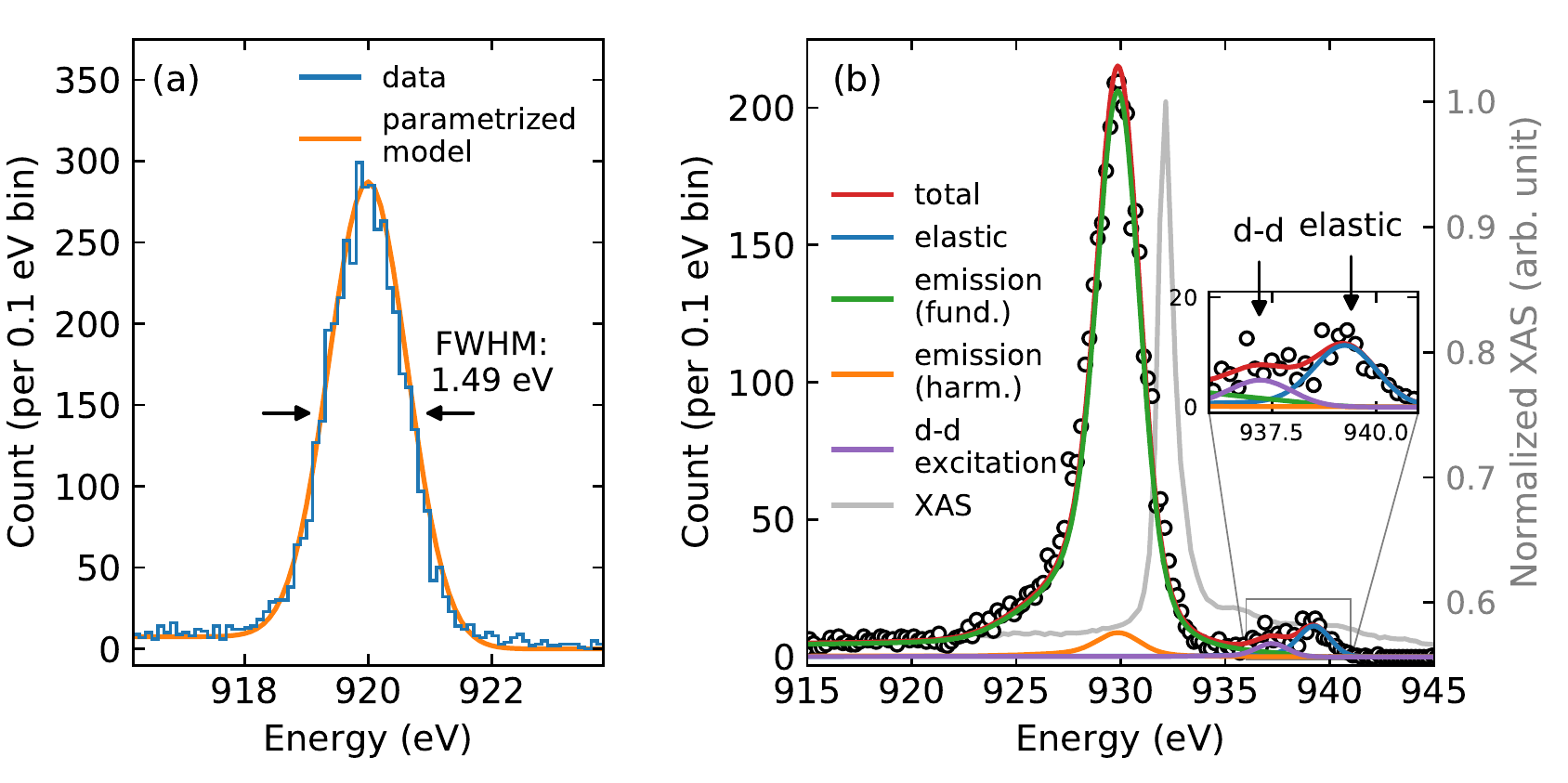}
    \caption{
(a) Resolution function of the TES (blue curve) determined from diffuse elastic scattering from a thin layer of gold on polished silicon at a fixed incident beam energy of 920 eV. (orange curve) Fit function used in analysis of the data. (b) LBCO spectrum taken with sensor \#15 far from the CDW Bragg condition showing elastic, $d$-$d$, and fluorescence emission features in the data. The colored lines represent the different components of the fit model (see text). The spectrum is measured at a fixed incident beam energy of 938.9 eV. Cu $L_{3/2}$ XAS spectrum of LBCO showing the absorption maximum. The XAS curve is plotted against
incident photon energy while the other curves are ploitted against scattered photon energy.}
    \label{fig:rixs_plots}
\end{figure}

A representative TES spectrum illustrating the basic features of the data is shown in Fig.~2(b). This histogram shows the number of photon events detected by sensor \verb+#+15 at an incident beam energy of 938.9 eV for an integration time of 2 min, as a function of detected photon energy. For comparison, plotted together is the Cu $L_{3/2}$ x-ray absorption spectrum (XAS) taken in total electron yield mode from the LBCO sample. This spectrum is plotted against incident photon energy and shows that the absorption maximum occurs at 932.2 eV. 

While the 1.5 eV energy resolution is modest, the expected features in the spectra are clearly visible, including the elastic scattering, $d$-$d$ excitations which are offset from the elastic peak by about 2 eV, and an incoherent emission peak at 930 eV that dominates the overall signal. The $d$-$d$ excitations are actually a coherent inelastic scattering effect, making Fig.~2(b) an example of a valence band resonant inelastic x-ray scattering (RIXS) spectrum taken using the intrinsic energy resolution of a solid state detector. Note that the incident beam is highly detuned from the absorption maximum at 932.2 eV. That the $d$-$d$ excitations are nevertheless visible is a demonstration of the high sensitivity of the TES detector. 

\begin{figure*}
    \centering
    \includegraphics[width=\textwidth]{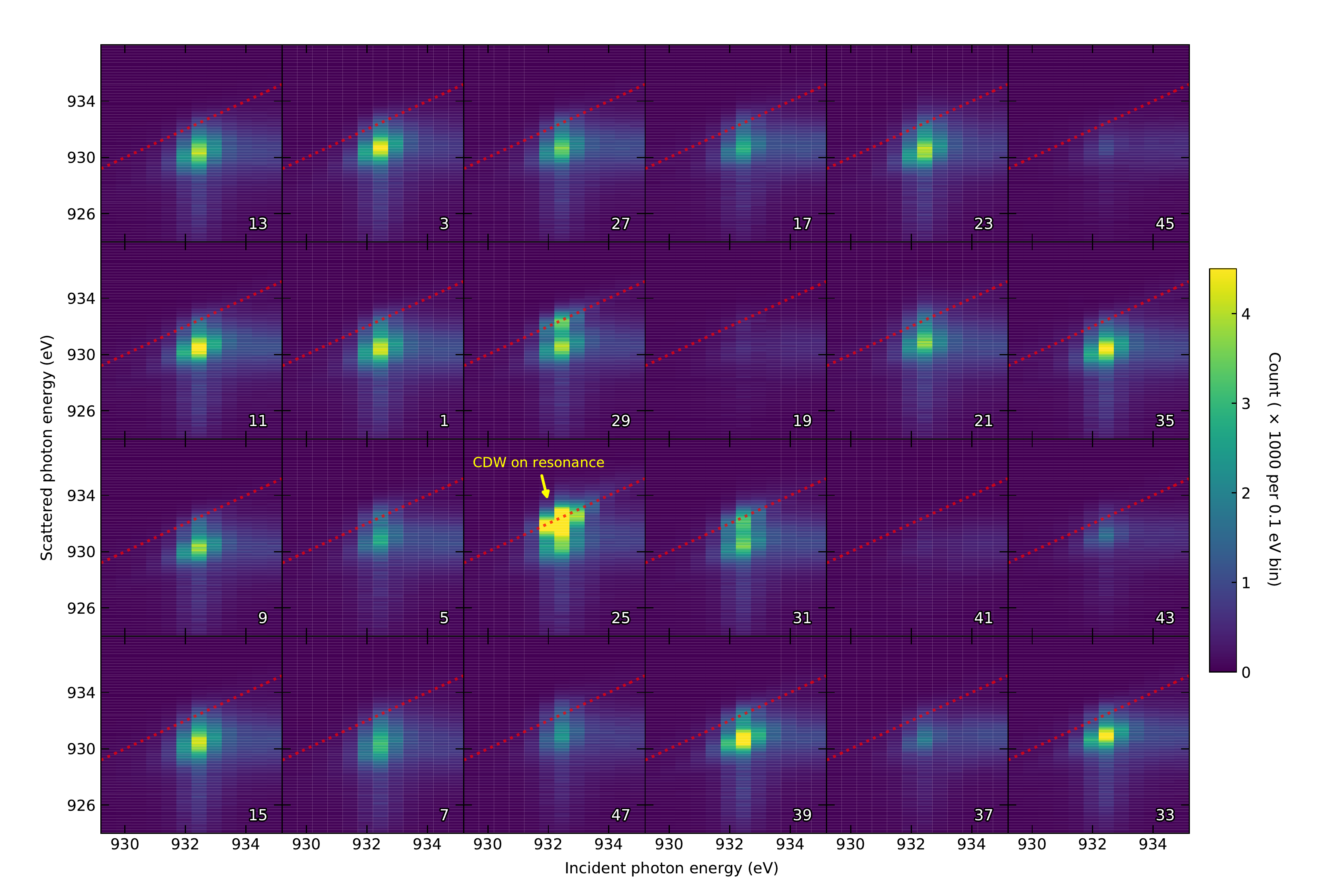}
    \caption{
Incident energy dependence of the scattered intensity from LBCO as measured by the 24 active sensors in the TES array. The panels are arranged to roughly correspond to the sensor layout shown in Fig.~1(b).
The incident photon energy was tuned from 929.2 eV to 935.8 eV in 0.5 eV steps. The sample angle was changed from $35.1^\circ$ to $35.3^\circ$, in coordination with the incident beam energy, to keep the in-plane component of the momentum transfer fixed during the scan. 
The CDW signal is observed in sensor \#25 whose location corresponds to a momentum vector $\mathbf{q} = (0.23,0,1.54)$. The red lines indicate the location of the elastic scattering ($\omega_i=\omega_s$).}
\end{figure*}

The incident energy dependence of the CDW scattering, as observed with the TES array, is summarized in Fig.~3, which shows the measured intensity as a function of both incident and detected photon energy on a color plot for each of the 24 active sensors. The sample angle was adjusted in coordination with the incident energy to keep the in-plane component of the momentum transfer fixed, with sensor \#25 aligned to a momentum of (0.23, 0, 1.54) to detect diffraction from the CDW. The integration time for each distinct incident energy was 5 min. The incident photon energy was stepped in 0.5 eV increments through the Cu $L_{3/2}$ edge at 932.2 eV, which corresponds to electric-dipole transitions into the $3d$ shell, enhancing sensitivity to valence band order \cite{abbamonte2005}. 

Fig.~3 demonstrates a few important general features of resonant scattering that illustrate how to best use a TES detector for RSXS experiments. Below resonance, photoabsorption from Cu $2p$ $\rightarrow$ $3d$ transitions is not allowed. However, resonant scattering--both elastic and inelastic--can still take place. The scattering will be ``off-shell," meaning whatever energy denominators are present in the cross section are not perfectly minimized, but resonant scattering can still take place albeit with somewhat reduced intensity. The elastic scattering, which is the signal of interest here, shows up on the diagonals in each panel in Fig.~3 (red lines). When the incident beam energy is above the resonance, the Cu$^{2+}$ fluorescence line becomes strong and sits at fixed emitted photon energy \cite{Eisenberger1976}. At the incident x-ray beam energy of 932.2 eV, which corresponds to the resonance maximum, the intensity of both the elastic scattering and the x-ray fluorescence is maximized. However, the different spectral components are intermixed and cannot be resolved using the 1.5 eV resolution of the TES. Optimal background rejection therefore requires slight detuning of the energy from the resonance. For this reason, studies of the charge order were carried out at an incident beam energy of 933.7 eV at which the scattering retains a considerable amount of the resonant enhancement and the elastic component is energetically resolvable using the resolution of the TES.

\begin{figure*}
    \centering
    \includegraphics[width=\textwidth]{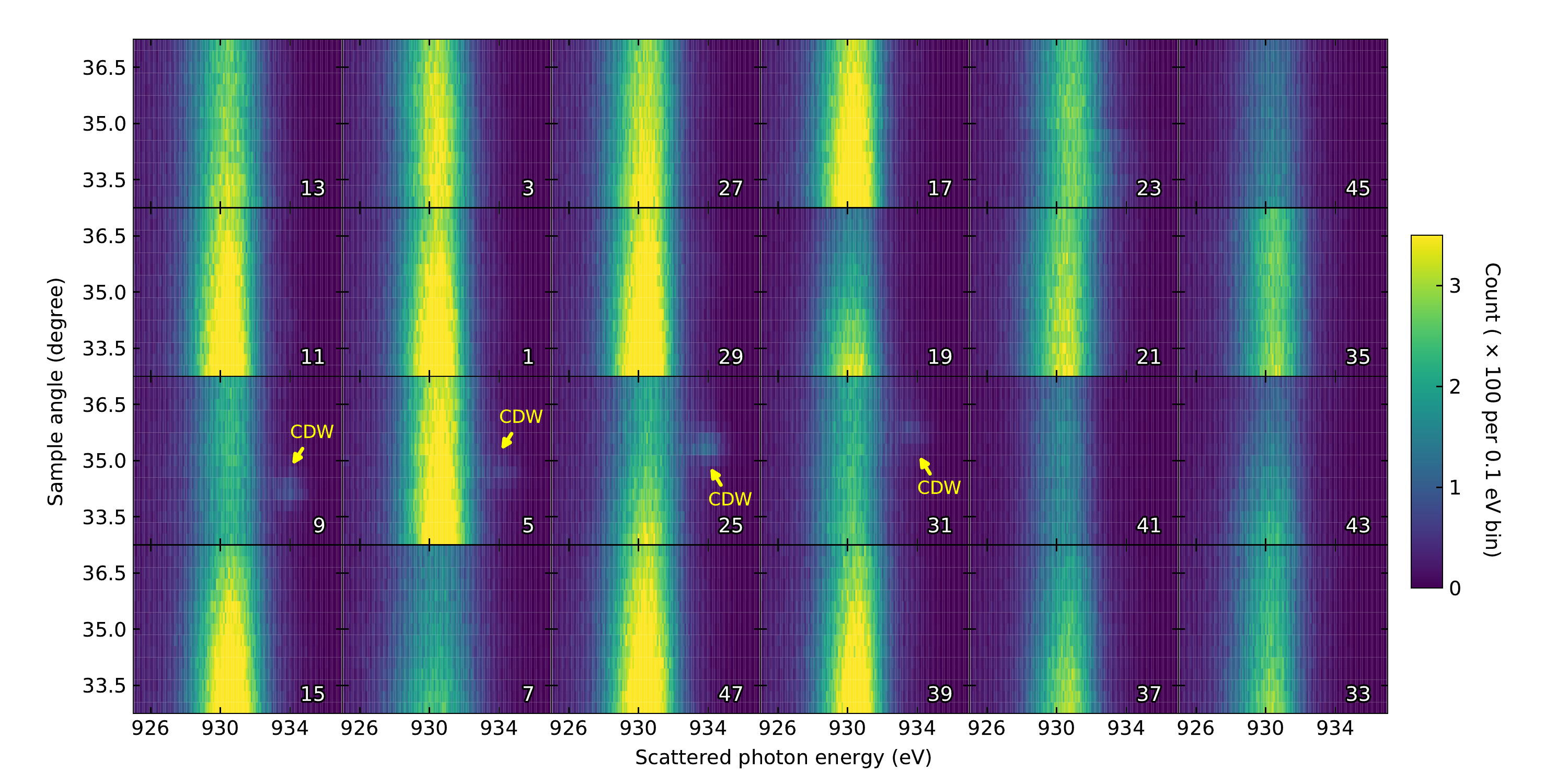}
    \caption{
Sample angle dependence of the scattered intensity from LBCO. This data set is essentially a scan of the momentum transfer vector of each of the 24 active sensors of the TES detector. The panels are numbered according to the naming convention in Fig.~1(b). The CDW is visible as a peak at the scattered energy of 933.7 eV which corresponds to the incident beam energy. The CDW sweeps across sensors 9, 5, 25, and 31. 
}
\end{figure*}

Having chosen an incident beam energy, we performed momentum scans through the charge order peak by rotating the $\theta$ angle of the sample, in 0.1$^\circ$ increments, at fixed incident energy (Fig.~4). In this scan the CDW sweeps across sensors \verb+#+9,  \verb+#+5,  \verb+#+25, and \verb+#+31, in which it is clearly visible as a peak at 933.7 eV that is resolvable from the fluorescence line at 930.1 eV. The data in Fig.~4 contain enough information to determine a quantitative, energy-resolved elastic lineshape of the CDW, which should be superior to what was possible in early studies \cite{abbamonte2005}. However, doing so requires some means to extract the elastic scattering from the individual spectra. 

\subsection{SPECTRAL DECOMPOSITION}

Having measured energy-resolved spectra in the vicinity of the CDW, a means to extract the coherent elastic signal from the data is needed. The individual spectra (Fig.~2(b)) consist of three main features: (1) the elastic line itself, which is resolution-limited and always resides in the spectrum at the incident beam energy, (2) the $d$-$d$ excitations, which reside $\sim$2 eV below the elastic line and are also resolution limited, and (3) the fluorescence line, which is the primary source of background. As shown in Fig.~3, the behavior of the fluorescence feature is complicated, evolving from a Raman-like excitation to an incoherent emission feature at 930 eV as the beam energy is tuned through the resonance \cite{Eisenberger1976}. A detailed spectroscopic model of these features is needed to quantify the amount of elastic scattering in the spectra. This model should properly account for the finite detector resolution, which is similar to the separation between the elastic and fluorescence features in most of the data, by convolving with the response curve in Fig.~2(a). 

The elastic scattering and $d$-$d$ excitations are both resolution-limited. We therefore model both as Dirac delta functions, which will broaden into replicas of the resolution function of the instrument when convolved with the response curve (Fig.~2(a)). 

To model the fluorescence feature, we apply a modified version of the Eisenberger-Platzman-Winick (EPW) model of x-ray emission \cite{Eisenberger1976}. Taking only the resonant terms in the Kramers-Heisenberg equation and assuming constant matrix elements and density of states in the continuum, the differential cross-section has the form

\begin{widetext}
\begin{equation}
\frac{d\sigma}{d\omega_2}(\omega_1, \omega_2) = \int_0^{\infty}d\epsilon_1\int_0^{\infty}d\epsilon_2 \frac{(1 - n_k(\epsilon_1))n_k(\epsilon_2)}{(\omega_1 - \epsilon_1 - \Omega_L)^2 + \Gamma_L^2} \delta(\epsilon_1 - \epsilon_2 - \omega_1 + \omega_2),
\end{equation}
\end{widetext}

\noindent
where $\epsilon_1$ and $\epsilon_2$ are respectively the energies of the photoexcited electron and decaying hole measured with respect to the bottom of the valence band, $\omega_1$ and $\omega_2$ are respectively the energies of incoming and outgoing x-ray photons, $\Omega_L$ is the energy separation between Cu L$_{3/2}$ core level and the bottom of the valence band, $\Gamma_L$ is the lifetime of the core-hole, and $n_k(\epsilon)$ is the Fermi function. Assuming temperature effects are negligible we take $n_k (\epsilon) = \theta(E_F- \epsilon)$, where $\theta$ is the Heaviside function and $E_F$ is the Fermi energy, in which case the cross-section reduces to

\begin{widetext}
\begin{equation}
\frac{d\sigma}{d\omega_2} =
\begin{cases}
0 &\text{if } \omega_2 < \omega_1 \\
\frac{1}{\Gamma_L} \left[ \arctan\left(\frac{E_F + \Omega_L - \omega_2}{\Gamma_L}\right) - \arctan\left(\frac{E_F + \Omega_L - \omega_1}{\Gamma_L}\right) \right] &\text{if } \omega_1 - E_F < \omega_2 < \omega_1 \\
\frac{1}{\Gamma_L} \left[ \arctan\left(\frac{E_F + \Omega_L - \omega_2}{\Gamma_L}\right) - \arctan\left(\frac{\Omega_L - \omega_2}{\Gamma_L}\right) \right] &\text{if } \omega_2 < \omega_1 - E_F
\end{cases}.
\end{equation}
\end{widetext}

A least-squares fit of this model to a sample spectrum, properly convolving the resolution function of the detector, is shown in Fig.~2(b). The intensity of the fluorescence emission is proportional to the differential cross-section \cite{fink2013}, which is given by Eq.~3 (Fig.~2(b), green curve). The quality of the fit is reasonable and accounts for all the primary spectral features. A small amount of fluorescence is present even when the x-ray energy is below the edge because of second harmonic contamination in the x-ray beam, i.e., photons with double the nominal energy. This extra component was explicitly included in the fit (Fig.~2(b), orange curve). 

\begin{figure}
    \centering
    \includegraphics[width=0.495\textwidth]{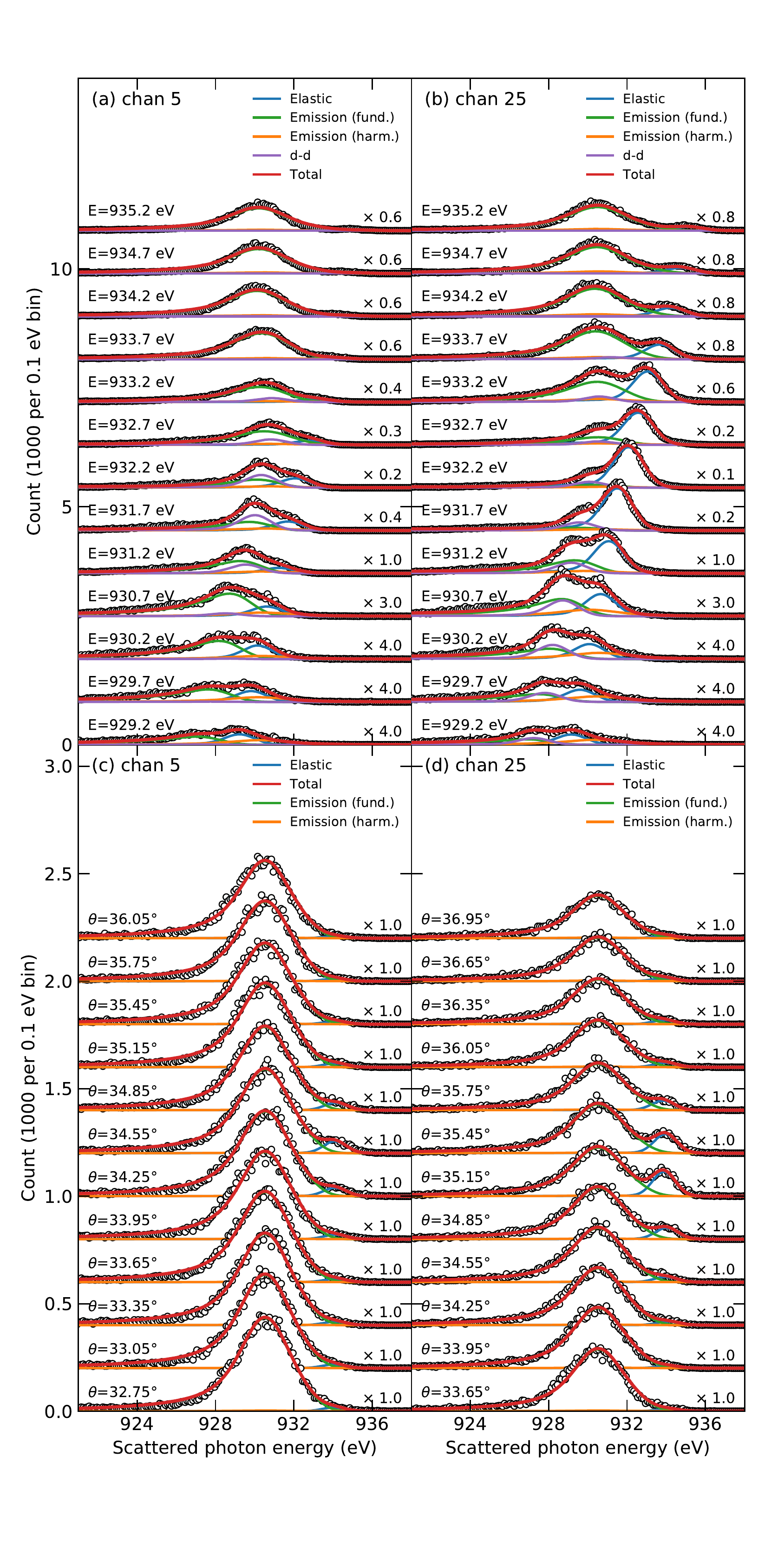}
    \caption{
Application of the fit model to the TES spectra from sensors \#5 and \#25. Each curve represents a cut as a function of scattered photon energy through panels \#5 and \#25 in Figs. 3 and 4. Some of the spectra have been scaled for visibility. 
(a) Individual TES spectra for sensor \#5 for various incident beam energies, from Fig.~3. Solid lines show different terms in the fit model and open circles are the experimental data points. (b) Same plot as (a) for sensor \#25, where the CDW scattering is strongest. At each incident beam energy the sample angle is changed in a coordinated way to keep the momentum transfer vector fixed. (c) Individual TES spectra for sensor \#5 for various sample angles, showing enhanced elastic CDW scattering for $\theta = 34.55^\circ$, from Fig.~4. (d) Same plot as (c) for sensor \#25, showing enhanced elastic CDW scattering at $\theta = 35.15^\circ$. The angle-dependent spectra were taken at a fixed incident energy of 933.7 eV.
}
    \label{fig:spectra_chan5_25}
\end{figure}

We now use this model to extract the elastic counts from sensors \verb+#+25 and \verb+#+5, which correspond to the integrated intensity of the elastic feature (Fig.~5) at the peak of the CDW reflection and a location on its tail, respectively. The result of the fits to the individual curves (Figs. 3,4) is shown in line plots in Fig.~5. The model allows each spectrum to be decomposed into discrete components, the counts being assigned to bins corresponding to $d$-$d$, fluorescence, and the elastic counts of interest. 

We are keenly interested in how the statistical quality of RSXS data is improved by introducing energy resolution into the measurement. The two values of interest are the elastic counts, $N_\text{elastic}$, and the total photon counts, $N_\text{total}$. $N_\text{total}$ is the energy-integrated total counts collected by a TES sensor, and represents  
how a given measurement would look had it been done with an energy-integrating detector of the same quantum efficiency under the same conditions. 

We assigned error values to the elastic counts in each spectrum by taking the square root of the corresponding diagonal value of the variance-covariance matrix of fit parameters \cite{taylorbook}. This error value obtained in this manner, $\Delta N_\text{elastic}$, accounts for statistical error as well as systematic error resulting from imperfection of the fit model. If the model were perfect, the error would reduce to the Poisson result, $\Delta N_\text{elastic}=\sqrt{N_\text{elastic}}$. In practice, we expect the error to exceed the Poisson value since the model will not fit the spectrum perfectly. The error in the energy-integrated counts is determined purely by Poisson counting statistics, $\Delta N_\text{total}=\sqrt{N_\text{total}}$.

The attribute of a TES detector that provides a performance increase for RSXS measurements is its ability to separate out the elastic counts, $N_\text{elastic}$, from all the photon events, $N_\text{total} = N_\text{elastic} + N_\text{background}$. The noise level, which in all cases is limited by Poisson counting statistics, is $\sqrt{N_\text{elastic}}$ for the former quantity and $\sqrt{N_\text{total}}$ for the latter. The TES therefore provides a significant reduction in noise and improvement in data quality, even in cases where the average background is known. The TES is particularly effective when the elastic signal is weak and the background level is high.

The results of the fits in Fig.~5 are summarized Fig.~6, which shows the elastic counts, $N_\text{elastic}$, for sensors \verb+#+5 and \verb+#+25 plotted against energy and angle. The error bars in Fig.~6 represent $\pm \Delta N_\text{elastic}$. For comparison, on the same plots we show the energy-integrated counts, $N_\text{total}$, with the error bars corresponding to $\pm \Delta N_\text{total}=\pm \sqrt{N_\text{total}}$.

%We are keenly interested in how the statistical quality of RSXS data is improved by introducing energy resolution into the measurement. We therefore assigned error values to the elastic counts in each spectrum, $N_\text{elastic}$, by taking the square root of the corresponding diagonal value of the variance-covariance matrix of fit parameters \cite{taylorbook}. This error value obtained in this manner, $\Delta N_\text{elastic}$, accounts for statistical error as well as systematic error resulting from imperfection of the fit model. If the model were perfect the error would reduce to the Poisson result, $\Delta N_\text{elastic}=\sqrt{N_\text{elastic}}$. In practice, we expect the error to exceed the Poisson value since the model will not fit the spectrum perfectly. 

%The attribute of a TES detector that provides a performance increase for RSXS measurements is its ability to separate out the elastic counts, $N_\text{elastic}$, from all the photon events, $N_\text{total} = N_\text{elastic} + N_\text{background}$. The noise level, which in all cases is limited by Poisson counting statistics, is $\sqrt{N_\text{elastic}}$ for the former quantity and $\sqrt{N_\text{total}}$ for the latter. The TES therefore provides a significant reduction in noise and improvement in data quality, even in cases where the average background is known. The TES is particularly effective when the elastic signal is weak and the background level high. 

\begin{figure}
    \centering
    \includegraphics[width=0.48\textwidth]{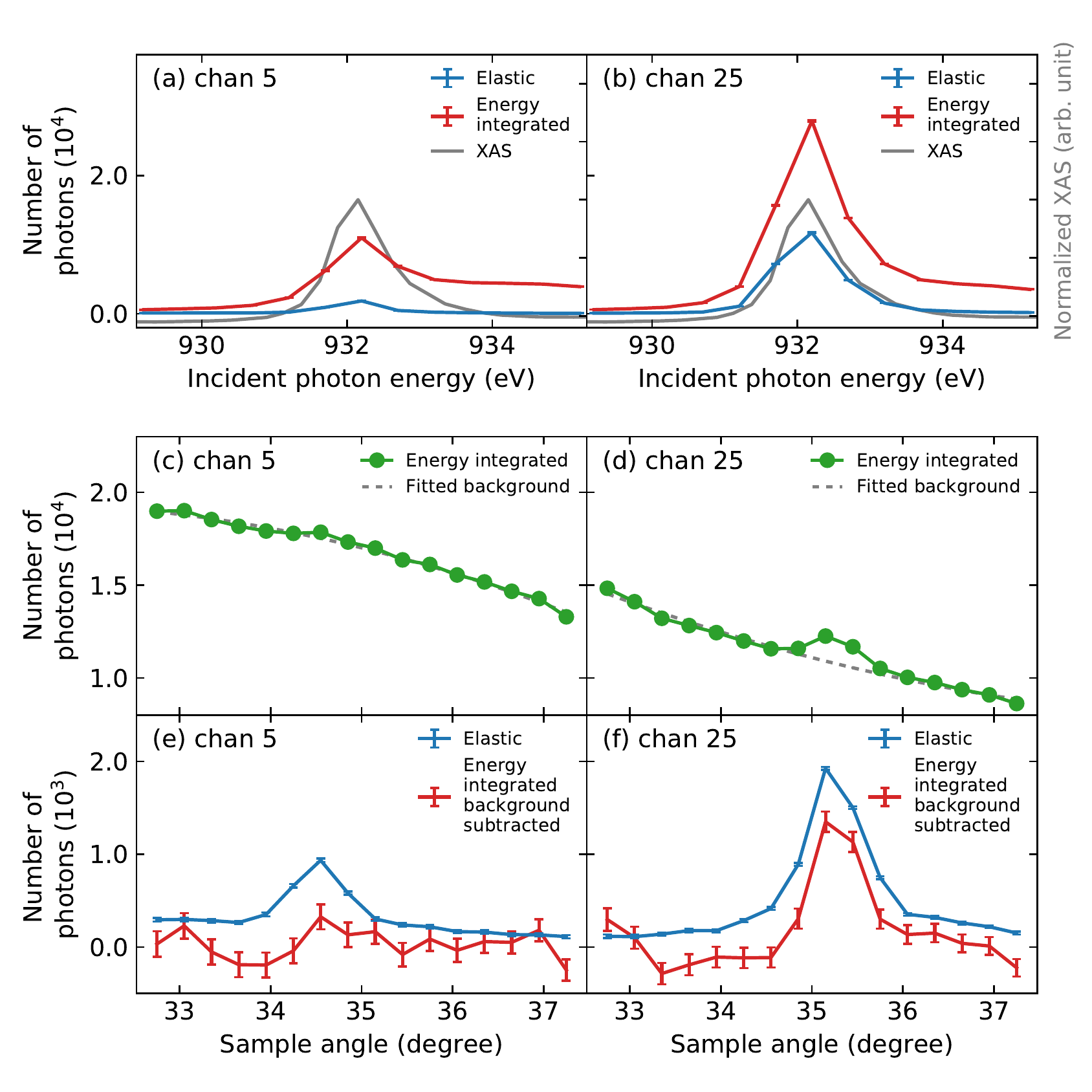}
    \caption{
Incident energy and angle dependence of the elastic scattering, as determined by the data fits in Fig.~5, compared to what would have been measured with an equivalent, energy-integrating detector (energy-integrated curves have been scaled to allow visual comparison). (a) Energy dependence of the elastic CDW scattering in sensor \#5 (blue) compared to the energy-integrated result (red). The energy-integrated curve exhibits an asymmetry and edge jump due to significant contribution from background fluorescence. The XAS spectrum, also shown in Fig. 2(b), is displayed for comparison. (b) Same plot as panel (a) for sensor \#25. (c) Angle-dependence of the energy-integrated CDW scattering in sensor \#5. The CDW peak is barely visible above the fluorescence background. (d) Same plot as panel (c) for sensor \#25. (e) Angle-dependence of the elastic CDW scattering in sensor \#5 (blue) compared to the energy-integrated result (red). The background was removed from the energy-integrated data by subtracting a second order polynomial fit, allowing a direct visual comparison between the points. The CDW feature is much more clearly revealed in the elastic scattering data. It's angular width of 1$^\circ$ is consistent with prior studies \cite{abbamonte2005} (f) Same plot as panel (e) for sensor \#25. 
}
    \label{fig:rsxs_summary}
\end{figure}

%The results of the fits in Fig.~5 are summarized Fig.~6, which shows the elastic counts, $N_\text{elastic}$, for sensors \verb+#+5 and \verb+#+25 plotted against energy and angle. The error bars in Fig.~6 represent $\pm \Delta N_\text{elastic}$. For comparison, on the same plots we show the energy-integrated counts, $N_\text{total}$, which represents the total photon counts on the corresponding TES sensor. $N_\text{total}$ represents how the spectrum would have looked had it been measured with an energy-integrating detector of the same quantum efficiency under the same conditions.

%The error in the energy-integrated points is determined purely by Poisson counting statistics, so the error bars correspond to $\pm \sqrt{N_\text{total}}$.

Focusing first on the energy spectrum from sensor \verb+#+25 (Fig.~6(b)), the elastic curve differs from the energy-integrated curve in that the edge jump, seen as an increased intensity for energies greater than about 934 eV, is absent. This indicates that there is enhanced fluorescence background when the energy is tuned far above the edge, but there is little enhancement of the resonant elastic scattering there. In other words, the apparent jump is entirely due to background fluorescence, and is not a property of the RSXS signal itself. 

A dramatic difference is seen in the angle scans through the CDW peak. Had the measurement been done with an energy-integrating detector, it would have resembled Fig.~6(d), which shows a CDW peak with a background similar to past studies \cite{abbamonte2005}. Using the fit model to extract the elastic scattering results in the curve shown in Fig.~6(f), which is plotted next to the energy-integrated spectrum with a background line subtracted. The curves in Fig.~6(f) differ in two important respects. First, the noise level in the elastic curve is lower. Second, the elastic curve does not go to zero when the crystal is rotated far from the Bragg condition (see, for example, the value at $\theta = 33^\circ$). This indicates that the elastic scattering peak has long tails, which cannot be detected without an energy-resolving detector. These tails could be a property of the CDW, or they could arise from some other effect, such as a rough sample surface. Using an energy-resolving TES detector opens up the possibility of studying these tails and determining their physical origin. 

We turn now to sensor \verb+#+5, which only detects a distant tail of the CDW so is a good test case for how the TES performs when applied to weak signals. The result of the energy spectrum (Fig.~6(a)) is similar to sensor \verb+#+25, with the edge jump being eliminated. The angle scans, however, show a qualitative difference: While the CDW peak is hardly visible above the noise level in the energy-integrated curve (Figs. 6(c), (e)), the elastic spectrum shows a well-defined peak with a width similar to what was observed in sensor \verb+#+25. In this case the ability to see the CDW peak at all requires an energy-resolved detector. Evidently the performance advantages of a TES are particularly significant for weak signals that are at or below the noise level of the fluorescence background. 

\subsection{DISCUSSION}

We are now in a position to make a quantitative statement about the advantage of introducing modest energy resolution into RSXS experiments. We focus on four data points in Fig.~6 corresponding to the peak in sensor \verb+#+5 ($\theta=34.55^\circ$) (Fig.~6(e)), a location on the tail in the same curve ($\theta=36.35^\circ$), the peak in sensor \verb+#+25 ($\theta=35.15^\circ$) (Fig.~6(f)), and a location in its tail ($\theta=36.35^\circ$). At these points we compare the magnitudes of the elastic counts, $N_\text{elastic}$, the energy-integrated counts, $N_\text{total}$, and the error values on these quantities, $\Delta N_\text{elastic}$ and $\Delta N_\text{total}$, as summarized in Table I. $N_\text{total}$ in this table represents the value after having subtracted a background line, as illustrated in Fig.~6(e),(f). 

\begin{table}
\caption{Summary of the numbers of elastic and energy-integrated counts at the peak and in the tail of the CDW in LBCO for sensors 5 and 25.}
\begin{ruledtabular}
\begin{tabular}{c|c|c|c}
${\bf \theta}$&{\bf Sensor}&{\bf $N_\text{total} \pm \Delta N_\text{total}$}&{\bf $N_\text{elastic} \pm \Delta N_\text{elastic}$}\\
\hline
$34.55^\circ$ (peak)&5&323 $\pm$ 133&883 $\pm$ 36\\
$35.15^\circ$ (peak)&25&1336 $\pm$ 110&1879 $\pm$ 25\\
$36.35^\circ$ (tail)&5&58 $\pm$ 122&122 $\pm$ 37\\
$36.35^\circ$ (tail)&25&148 $\pm$ 98&290 $\pm$ 23\\
\end{tabular}
\end{ruledtabular}
\end{table}

Note that some of the error values in Table I are larger than $\sqrt{N}$. In the case of the total counts, $N_\text{total}$, this is because the fluorescence background also contributes to the noise level. In the case of the elastic counts, $N_\text{elastic}$, an increased value of the error can be expected if the fit function is imperfect (Fig.~5).

We begin our comparison with the peak count rate in sensor \verb+#+25. A useful figure of merit is the quality ratio, $N/\Delta N$, which would reduce to the signal-to-noise ratio in cases where $\Delta N$ is limited by counting statistics. For the energy-integrated measurement, $N_\text{total}/\Delta N_\text{total}$=12.1. For the elastic case, this ratio $N_\text{elastic}/\Delta N_\text{elastic}$=75.2. We conclude that using energy analysis provides a factor of 6 improvement in data quality in this case.

Turning now to the peak in sensor \verb+#+5, the relevant ratios are $N_\text{total}/\Delta N_\text{total}$=2.4 and $N_\text{elastic}/\Delta N_\text{elastic}$=24.5. In this case adding energy resolution results in a factor of 10 improvement in signal quality. This seems consistent with the qualitative observation that, while a peak is hardly visible in the energy-integrated measurement, it is clearly visible in the energy-resolved curve (Fig.~6(e)). This result illustrates an important point about the applicability of TES detectors in scattering: the performance improvement is higher when studying weak signals that lie at or below the noise level from the fluorescence background. 

This point is further validated by examining the tails of the peak. In the case of sensor \verb+#+25, the ratios are $N_\text{total}/\Delta N_\text{total}$=1.5 and $N_\text{elastic}/\Delta N_\text{elastic}$=12.6, implying a performance enhancement of 8.4. For the case of \verb+#+5, we have $N_\text{total}/\Delta N_\text{total}$=0.47 and $N_\text{elastic}/\Delta N_\text{elastic}$=3.2, implying a performance improvement of 7.0.

 These numbers represent the enhancement in performance {\it purely} from the introduction of moderate energy resolution in RSXS studies. An explicit performance comparison between a TES and a specific detector would also require accounting for quantum efficiency, solid angle coverage, number of pixels or sensors in the device, etc. Nevertheless, it is clear that, all other things being equal, introducing even modest energy resolution can improve RSXS studies by an order of magnitude. 

In conclusion, we have evaluated the performance of a TES detector with 1.5 eV energy resolution for RSXS measurements from the prototypical stripe-ordered CDW material, La$_{2-x}$Ba$_x$CuO$_4$. We have found that incorporating modest energy resolution provides a factor of 5 to 10 times improvement in performance compared to equivalent, energy-integrated measurements. TES technology is likely to lead to significant new discoveries in studies of materials with emergent valence band instabilities, particularly those exhibiting glassy or short-ranged order, in which the scattering is diffuse and may lie near or below the noise level of the fluorescence background. 

\begin{acknowledgments}
We acknowledge Jun-Sik Lee and Sang Jun Lee of SLAC for helpful discussions, and G. J. MacDougall for pointing out an error in the data analysis. RSXS experiments were supported by DOE BES grant
no. DE-FG02-06ER46285. Design, construction, and operation of the TES detector was supported by the NIST Innovations in Measurement Science program. Use of the Advanced Photon Source was supported by DOE BES grant no. DEAC02-06CH11357, where construction of the endstations was supported by NSF grant no. DMR-0703406. K. M. acknowledges support from a National Research Council Postdoctoral Fellowship.
\end{acknowledgments}

\bibliography{research_v4}

\end{document}